\documentclass[useAMS,usenatbib,usegraphicx,referee]{mn2e}

\title{Modeling anomalous extinction using nanodiamonds}

\author[Rakesh K Rai and Shantanu Rastogi]{Rakesh K Rai and Shantanu Rastogi\thanks{E-mail: shantanu\_r@hotmail.com}\\
Department of Physics, D.D.U. Gorakhpur University, Gorakhpur - 273009, India}

\begin{document}

\maketitle

\label{firstpage}

\begin{abstract}
Modeling extinction along anomalous/non-CCM sightlines, which are characterized by a broad $217.5~nm$ bump and steep far-ultraviolet (FUV) rise, is reported. The extinction along these sightlines, viz. {HD 210121}, {HD 204827}, {HD 29647} and {HD 62542}, is difficult to reproduce using standard silicate and graphite grains. Very good match with the observed extinction is obtained by considering nanodiamond component as part of carbonaceous matter. Most of these sightlines are rich in carbon and are invariably backed by a young hot stellar object. Nanodiamond is taken as core within amorphous carbon and graphite. These core-mantle particles taken as additional components along with graphite and silicates lead to reduction in the silicate requirement. The abundance of carbonaceous matter is not affected as a very small fraction of nanodiamond is required. Extinction along sightlines that show steep FUV is also reported demonstrating the importance of nanodiamond component in all such regions.

\end{abstract}

\begin{keywords}
Interstellar Medium -- Nanodiamonds, Extinction.
\end{keywords}

\section{Introduction}

The extinction of stellar light through the Interstellar medium (ISM) follows the simple law that it varies linearly with wavelength inverse. Superposed on it are features like the $217.5~nm$ bump and the far-ultraviolet (FUV) rise \citep{fm86, fm88, whittet03}. Several attempts to parametrize the extinction curve along various sightlines include \citet{fm86} and $ R_V $ dependent general extinction formula by \citet{ccm89} (here after CCM). Most of the sightlines follow the CCM law, but there are a number of lines of sight for which the CCM law is not good enough \citep{cardelli88,clayton03a}. Certain regions show weak broad bump and strong FUV rise \citep{valencic03} that closely resemble extinction in Small Magellanic Cloud (SMC). These anomalous sightlines are referred to as non-CCM sightlines \citep{valencic03,valencic04} and include {HD 210121}, {HD 204827}, {HD 29647} and {HD 62542}.

The steep FUV rise in extinction is interpreted as the manifestation of high abundance of very small grains \citep{larson, clayton03b, valencic04}. These sightlines may have been exposed to shocks or strong UV radiation that disrupt large grains resulting in a size distribution favoring small grains \citep{valencic04}. Small $R_V$ for these objects point to smaller particles but, contrary to observation, will simultaneously lead to stronger $217.5~nm$ bump \citep{larson}. All the non-CCM sightlines contain a B type young star as background to dense clouds, e.g. {HD 210121} is a high latitude B3 V star and {HD 204827} is a B0 V star, and are background to dense clouds rich in carbon and its molecules \citep{roueff02,oka03}. 

Several attempts to model the non-CCM extinction have been made that include considering different grain compositions \citep{liGreenberg98}, different size distributions \citep{weingartner01} or both \citep{larson,clayton03b}. Considering the prevalence of smaller particles in harsh environments \citet{liGreenberg98} used the MRN distribution \citep{mrn77} in the size range $a \in [0.001,0.200]$ and $a \in [0.001,0.120]~ \mu m $, for silicates and graphite respectively, to model extinction along {HD 210121}. To explain the poor match of silicate-graphite model \citet{liGreenberg98} also attempted silicate-organic refractory core-mantle model that required 20\% more silicon than is available to condense into the solid phase. \citet{weingartner01} use separate silicate and graphite model with a new size distribution having higher concentration of smaller particles towards {HD 210121}. Considering clusters of grains \citet{rouleau97} show that for {HD 204827} the match is improved in the FUV but gets poor in the bump region.  Simultaneous matching of the bump and FUV rise is difficult.

These sightlines are in general carbon rich having traces of carbonaceous molecules \citep{roueff02,oka03,adamkovics05} hence the silicate fraction along these sightlines could be lower than the average galactic abundance. Incorporating other carbonaceous materials supposed to be present in ISM viz. organic refractory carbon \citep{liGreenberg97}, PAH \citep{draine07}, bucky onions \citep{chhowala03,li08}, carbon nanoparticles \citep{schnaiter98}, nanodiamonds \citep{kerk02,binette05,rai10} etc. seem more suitable to model extinction along these objects.

\citet{clayton03a} and \citet{sofia05} show good fits using three component silicate, graphite and amorphous carbon (AMC) models and applying maximum entropy method (MEM). These models use very low fraction of graphite and in a very narrow size range while the upper size cut-off for AMC is ($\sim 1 \mu m$), about 10 times larger. In the strong UV flux regions of the non-CCM sightlines graphite being more rugged to wear and tear is more probable. The destruction of AMC grains to smaller sizes is a more likely scenario \citep{cano-jones08}.

The $217.5~ nm $ bump has nearly a fixed position \citep{fm86}. This peak in graphite particles shifts towards longer wavelength on coating with different materials e.g. ice-coating \citep{draine-malhotra} and diamond coating \citep{mathis94}. Observations along {HD 29647} and {HD 62542} \citep{cardelli88} show an opposite trend i.e. a shift towards shorter wavelength that is accompanied by steep FUV rise. Nanodiamond (ND) extinction shows steep rise in FUV beyond $5.5~ \mu m^{-1}$ \citep{binette05,rai10} and considering ND as core inside graphite or AMC mantle lowers the bump and shifts the $217.5~ nm$ peak blue-ward \citep{rai08,rai10}. There is profile similarity of extinction incorporating NDs inside carbonaceous material with observations along non-CCM sightlines. Using ND core inside graphite mantle gives good match with the average galactic extinction particularly in the FUV \citep{rai10}. Therefore modeling of non-CCM sites and sites with steep FUV is attempted by incorporating NDs as additional component and is being reported.

\section{Component Materials in Modeling}

The observation of 3.43 and 3.53 $\mu m$ emission bands, attributed to hydrogenated NDs, show convincing presence of NDs in the ISM \citep{guillois99, kerk02}. On the basis of C--H stretch band observations \citet{goto09} suggest ND formation nearer to the star where there is stronger UV flux. NDs in meteorites are enveloped in AMC and organic compounds \citep{dai02} pointing to the possibility of ND formation near the Sun in pre-solar carbon already present in its environment.

Carbon can exist in different forms. Although graphite is the most stable allotrope and one of the most refractory material with sublimation point above 4000 K at one atmosphere \citep{c-book}, the different forms transform into each other under UV irradiation or high energy electron collisions \citep{banhart97,zaiser97}. Experimentally partial graphitization of ND due to heat \citep{leguilou07} or pressure conditions \citep{davydov07} has been reported. \citet{kwon07} theoretically show that surface graphitization of diamond under strong radiation field leads to core-mantle like structure. In-situ formation of nanodiamonds and transformation of graphite to diamond is also observed over a wide range of irradiation conditions \citep{daulton01}.

Simulating UV photolysis in dark clouds \citet{kouchi05} show that diamonds and graphite few nanometer in size can form in organic ice mixture subjected to UV irradiation. Both graphite and diamond could form directly within amorphous carbon. With evolution the amorphous component may transform into these forms or be removed altogether. ND and glassy carbon mixture, possible product of supernova shock-wave transformations, are detected in meteorites \citep{stroud11}. ND-AMC is likely to be the dominant component while ND-graphite may result from secondary processing or in harsher atmospheres.

Considering extinction using pure ND grains give an abrupt extreme FUV rise but also a peak/plateau $\sim$1400 \AA{} \citep{rai10} that is not observed along any sightline. The strong radiation field in non-CCM objects, which are generally young B type objects with carbon rich atmospheres, may lead to grain processing and formation of core-mantle type structures, with ND as core in graphite or AMC mantle. For the same reason large AMC grains, as considered by \citet{clayton03a, sofia05} in extinction models, are likely to be destroyed and easily re-cycled and re-formed \citep{cano-jones08}. Considering ND in carbonaceous core smoothens the extinction peak/plateau $\sim$1400 \AA{} \citep{rai10} and also explains the rare detection of features likely in pure ND.

\begin{figure*}
\centering
 \includegraphics{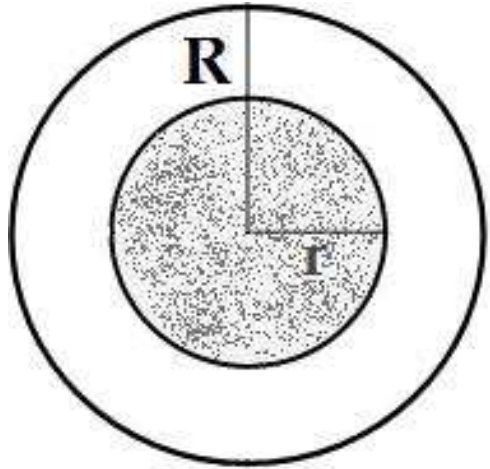}
\caption{Core-mantle spherical grains with nanodiamond of radius r}
\label{fig1}
\end{figure*}

In the present work core-mantle ND-graphite and/or ND-AMC (Figure \ref{fig1}) are treated as different materials and are used as separate components along with independent silicate and graphite components. NDs are structural changes in carbonaceous particles due to bond re-arrangement and hence size range and distribution for these core-mantle carbonaceous particles is assumed to be the same MRN distribution as that for graphite or AMC.

\textit{ND in Amorphous Carbon}: The AMC extinction has plateau/peak at 5.4 $\mu m^{-1}$ \citep{cecchi98}. This is not seen in observations. On considering ND core inside AMC mantle this feature modifies such that larger the ND core smaller is the plateau/peak \citep{rai08}. The ratio of ND core to AMC mantle radii is taken to be 3:4 considering the high possibility of AMC surface removal in rough atmospheres. The upper cut-off for ND radius is fixed at 0.010 $\mu m$ \citep{kerk02}. The highest ND volume is nearly 42 \% and decreases for particles of radii larger than 0.013 $\mu m$.

\textit{ND in Graphite}: The ND-graphite phase could result from secondary processing so the ratio of radii of ND core and graphite mantle is considered to be 1:2, smaller than in ND-AMC. The upper cut-off for ND radius is taken to be 0.005 $\mu m$. So the ratio of radii decreases continuously for graphite radii larger than 0.010 $\mu m$. Thus the highest ND volume of a single grain is 12.5 \% and is much smaller in the overall distribution.

\textit{Graphite and Silicates}: The graphite and silicate grains are considered to be spherical and in the size range $a \in [0.001,0.120]$ and $a \in [0.001,0.200]~ \mu m$ respectively, as used by \citet{liGreenberg98}, except in modeling {HD 210121} for which the smaller silicate grain size range $a \in [0.001,0.120]~ \mu m$ yields a better result.

\section{Calculations}

Extinction is obtained for 55 wavelengths. For each wavelength and for all the components, 50 or more size bins are considered in MRN distribution. The silicate and graphite optical data are taken from B.T. Draine's site\footnote{http:/www.astro.princeton.edu/$\sim$draine/dust/dust.diel.html}, optical data of AMC are from \citet{rouleau91} and the ND values are from \citet{mutschke04}. Mie theory for coated sphere is used in calculations of core-mantle particles \citep{bohren83}. 

The calculated values are fitted to reported observations by varying the fraction of grain components used. The $\chi^2 $ fitting applied here is defined by \citet{beving69} for wavelengths $i=1~to~ 55$, components $j=1~to~4$ and $f$ degrees of freedom:
\begin{displaymath}
\chi^{2}=\frac{\sum_i (\sum_j p_j.S^j_i-T_i)^2}{f}
\end{displaymath}
where 
\begin{displaymath}
S^j_i(\lambda_i)=[\frac{A(\lambda_i)-A(V)}{A(B)-A(V)}]_j
\end{displaymath}
is the normalized extinction curve of j\textit{th} grain component with $ p_j$  fraction and $T_i$ is the observed extinction at wavelength $\lambda_i$. The combination of different $p_j $ gives the minimized value of $\chi^2$. Since,
\begin{displaymath}
\sum_{j}p_j.S^j_i=\sum_{j}p_j.[\frac{A(\lambda_i)-A(V)}{A(B)-A(V)}]_j=\sum_{j}p_j.[\frac{A(\lambda_i)}{A(B)-A(V)}]_j-\sum_{j}p_j.[\frac{A(V)}{A(B)-A(V)}]_j
\end{displaymath}
therefore, for
\begin{displaymath}
\sum_{j}p_j.S^j_i=T_i=\frac{R_V}{A(V)}[\sum_{j}A^j(\lambda_i)]-R_V
\end{displaymath}
and using extinction for j\textit{th} component \citep{barbaro04}
\begin{displaymath}
A^j(\lambda_i)=1.086\pi N(H)\frac{n_{dj}}{n(H)}\int^{a_{max}}_{a_{min}} Q^B_{ext}a^{-1.5}da
\end{displaymath}
we obtain that abundance for j\textit{th} material is directly related to relative fraction $p_j $ by 
\begin{displaymath}
n_{d,j}= \frac{p_j.A_V}{1.086 \pi R_V N(H) A_j [\int^{a_{max}}_{a_{min}} Q^B_{ext}a^{-1.5}da-\int^{a_{max}}_{a_{min}} Q^V_{ext}a^{-1.5}da]}
\end{displaymath}

Thus knowing exactly the N(H) values, abundances for different components can be determined using a suitable multiplier in $p_j $. The best fit $p_j$ values therefore represent fractional abundance of various components. Simultaneously $R_V$ that depends on size distribution of the components used in modeling can also be theoretically determined

\begin{displaymath}
R_V=\sum_{j}p_j.[\frac{A(V)}{A(B)-A(V)}]_j
\end{displaymath}

\section{Results and Discussion}
The two component silicates and graphite model is modified with a third and in some cases a fourth component. ND core in AMC mantle and ND core in graphite being the two new components. Fitting is performed for non-CCM sightlines and sightlines having non-CCM type steep FUV rise. The data for observed extinction curve are taken from \citet{fm07}.

\begin{figure*}
\centering
 \includegraphics[width=165mm]{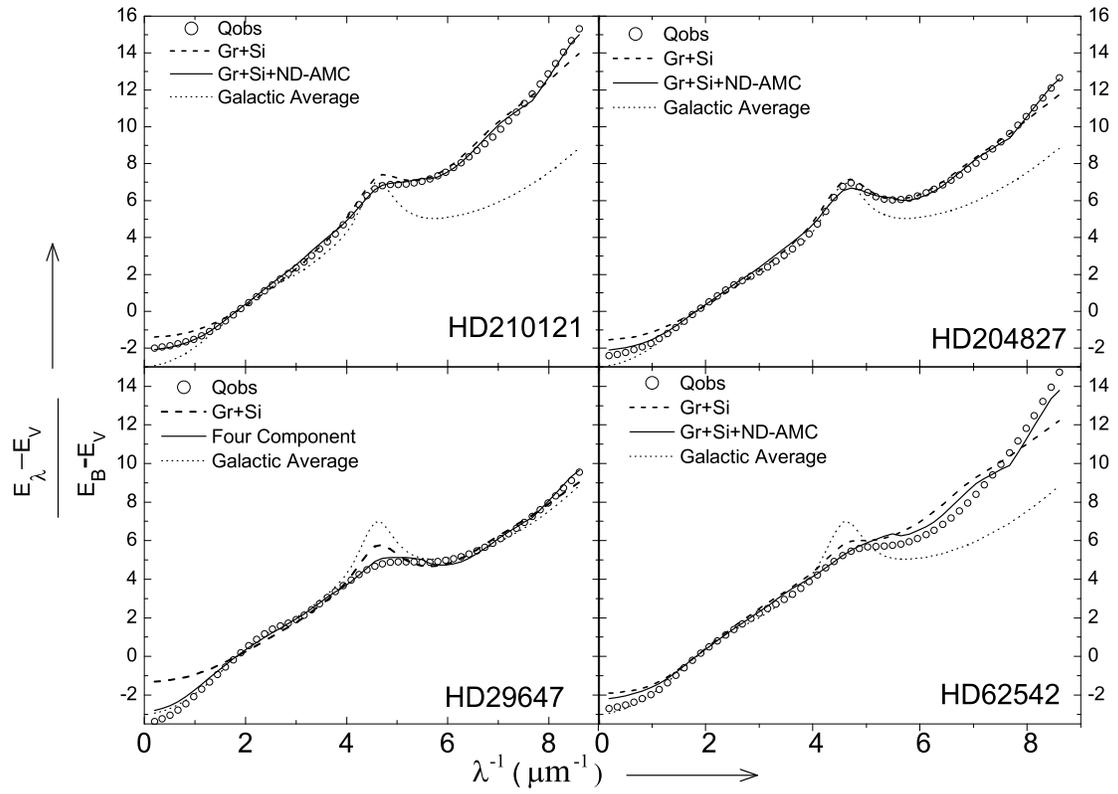}
\caption{Extinction curve modeling along non-CCM sightlines HD210121, HD 29647, HD 204827 and HD 62542}
\label{fig2}
\end{figure*}

All the models considered are presented in Table~\ref{tbl-1} for the four non-CCM objects. The $\chi^2$ value for the fitted curve and its corresponding theoretical $R_V$ are also given. The two component graphite-silicate model is presented for comparison (row 1). Replacing pure graphite component with ND-graphite core-mantle component (row 2) improves the fitting in all the cases except in {HD 204827}. Considering a three component model by addition of ND-AMC core-mantle component along with pure graphite and silicate components (row 3) gives a much better match for all the objects. Using only AMC as the third component improves fitting \citep{clayton03a, sofia05} but the FUV region is further improved in the present case with ND-AMC core-mantle component. The best fit curves are compared with observed values in Figure \ref{fig2} for the four sightlines, wherein the two component graphite-silicate model and the average galactic extinction are also shown for comparison.

Of the three parts of parametrization of \citet{fm86} the linear part is provided by silicates, Drude fit is due to graphite and ND-AMC addresses the FUV rise. This reduces considerably the requirement of silicates. For example in the case of {HD 210121} the silicate fraction goes down from 0.45 in the two component model to 0.18 in the model with ND-AMC as third component.

The line of sight {HD 210121}, obscured by high latitude translucent cloud, has a very small $R_V$ \citep{larson96} and has been a subject of several studies \citep{liGreenberg98,larson,weingartner01,clayton03a}. The three component graphite, silicate and ND-AMC model gives very good improvement from the graphite-silicate model in the infrared, the bump and the FUV regions (Figure \ref{fig2}). The calculated $R_V$ 2.14 is also close to the observed value 2.012. The grain size range considered in the present model is similar to earlier studies but the use of silicates is highly reduced. Most extinction models imply near solar or higher elemental abundances \citep{mathis96,weingartner01}. If ISM abundances are considered to be subsolar it becomes difficult to account for the overall extinction \citep{li05}. Though the elemental abundances and dust depletion data are likely to be constrained by observational and star formation effects \citep{li05}, incorporation of NDs comforts the need of silicates and the carbon budget is also not affected as very small fraction in diamond form is needed.

The sightline {HD 204827} is dense and rich in carbon molecules with material processed by supernova shocks \citep{valencic03}. Supernova annealing of carbonaceous material can lead to ND formation \citep{nuth92}. ND-AMC component is a viable dust component for this sightline and a good fit to the observational values is obtained. The silicate fraction is drastically reduced from 0.65 to 0.02.

{HD 29647} is a B7 V type highly reddened star in the Taurus dark cloud \citep{whittet04}. It has a very weak 217.5 $nm$ bump and steep FUV rise. In this region a four component model is needed that includes ND-graphite along with ND-AMC core mantle components, with a larger upper size cut-off (Table~\ref{tbl-1}, row 4). The molecular cloud TMC-1 is a rich repository of carbonaceous molecules with 26 first detection of molecules and ions in ISM\footnote{http://www.astrochymist.org/astrochymist\_ism.html} that include $HC_{11}N$ \citep{bell97}, the largest molecule, and $C_6H^-$ \citep{mccarthy06}, the first anion. The proximity of {HD 29647} to TMC-1 that is sampled along the line of sight \citep{whittet04} can induce complex structural changes in carbonaceous grains and both the core-mantle components could be present. The best fit is obtained with silicate fraction going to zero. Theoretical $R_V$ is still smaller (2.87) than the observed (3.456) and the match in the infrared region is a bit off. If the sampled medium is carbon rich cloud clump the silicate fraction should be quiet small, if not zero. By fixing silicate fraction to 0.02 the fitting $\chi^2$ is only slightly changed (Table~\ref{tbl-1}, row 5).

The non-CCM sightline {HD 62542}, which is a B3 V type star, has very weak bump and steep FUV. The object is situated at the ridge of expanding bubble of supernova remnant, which is heated by UV radiations of embedded stars and also some energy is supplied by shocked gases \citep{cardelli88}. The best $\chi^2$ value is 0.158 but the fitting beyond the bump is not as good as for other sights. Here again the silicate fraction obtained is zero, which when fixed to 0.02 has little effect on the model fitting (Table~\ref{tbl-1}, row 6). For such regions with active grain processing distributions other than MRN need to be considered. 

Extinction along {HD 29647} and {HD 62542} resemble extinction along SMC \citep{valencic03}. It was considered that silicon is rather undepleted in SMC \citep{welty01} but based on the Kramers-Kronig dispersion relation \citet{li06} show that silicates are important contributors to extinction in SMC. It may be interesting to consider ND component in the extinction models for SMC.

\begin{figure*}
\centering
 \includegraphics[width=165mm]{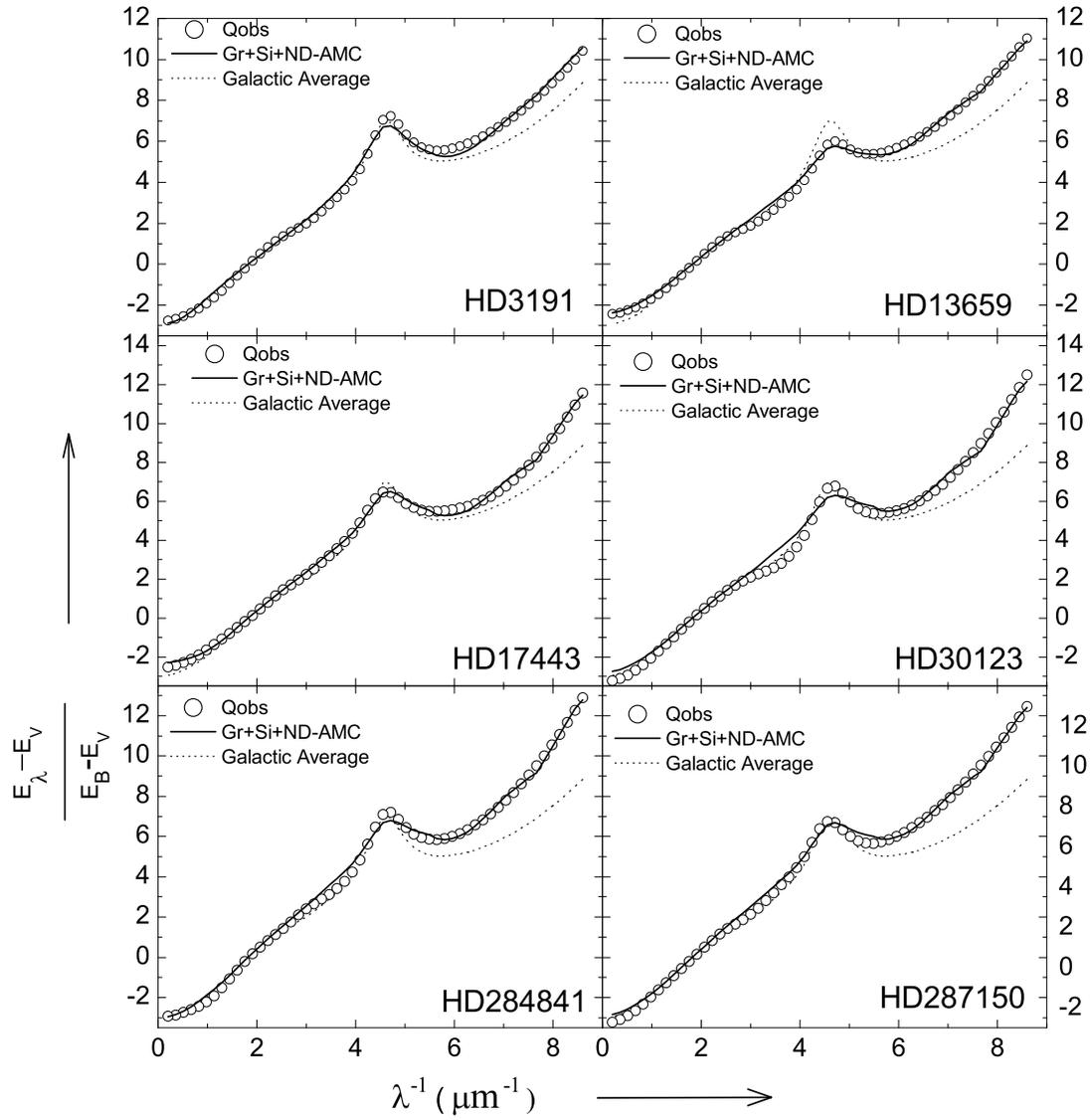}
\caption{Extinction curve modeling for stars with steep FUV rise}
\label{fig3}
\end{figure*}

The present model incorporating ND-AMC core-mantle additional component can also be extended to sightlines that resemble non-CCM extinction or have strong FUV extinction. Attempt is made to model few such sightlines i.e. {HD 3191}, {HD 13659}, {HD 17443}, {HD 30123}, {HD 284841} and {HD 287150}, and are presented in Table \ref{tbl-2} and Figure \ref{fig3}. For all these objects the bump feature is broadened \citep{fm07} and along all these lines of sight are young hot objects.

The sightline {HD 13659} shows type A anomalous extinction \citep{mazzei08} with a broad and weaker bump. {HD 30123} and {HD 284841} are in the Taurus open cluster {NGC 1647} and {HD 287150} is part of {NGC 1662} open cluster. Very good model fits are obtained with the $\chi^2$ values better than 0.05 for all objects except {HD 30123}. 

All the model fits require very small fraction of silicates. The lower size limit for silicate grains is taken to be $1~nm$ but the possibility of ultrasmall silicates ($a<1~nm$) cannot be ruled out \citep{liDraine01}. Ultrasmall silicates would increase the FUV excess without affecting the bump region. In the present models the required FUV excess is being taken care of by the ND component that simultaneously weakens and broadens the bump, which is required along these sightlines. Comparing the reported 10 objects it is further seen that the ND-graphite component was required only in {HD 29647}. This object is very close behind the Taurus cloud actively processing the grains thus consistent with the assumption that ND-graphite component results from secondary processing and in harsher environments.

The material in protoplanetary disks inherited from the ISM will contain ND mainly as ND-AMC component. The AMC mantle in the inner regions can be shredded by heating, shocks or grain shattering. This may leave the ND with dangling H atoms that can be responsible for the C--H stretch emission features from the inner regions \citep{goto09}.

It is interesting to note that along all non-CCM sightlines and sightlines with steep FUV extinction there is a young B type star processing the foreground ISM. {HD 29647} is processing the TMC-1 and {HD 283809} that samples the same cloud show non-CCM extinction \citep{valencic04}. In the same region the nearby B type star {HD 283800} is not sampling the cloud (\citet{whittet04}, fig.1) and shows normal extinction. The Orion molecular cloud {NGC 1976} has e.g. a young O type star {HD 37022} in its foreground \citep{shuping97} so the light reaching us does not sample/process the cloud and does not exhibit non-CCM extinction. Young hot objects in the vicinity of dense clouds induce grain processing and in carbonaceous regions can give rise to growth and transformations of different carbon metamorphs.

\section{CONCLUSION}

Experimental, theoretical and observational studies point to the possibility of carbon in diamond form in the ISM. Besides enhanced far-UV extinction \citep{binette05, rai10}, luminescence from diamond nano-crystals could be responsible for the Extended Red Emission (ERE) \citep{chang06}. 
Considering ND grains as core inside other carbonaceous matter \citep{rai10} show that anomalous sightlines having sharp FUV rise and weak/broad bump feature can be modeled by incorporating ND as a component.

The non-CCM sightlines have steep FUV rise and broad and sometimes weak $217.5~nm$ bump. The carbonaceous material along these sightlines can be processed leading to core-mantle type ND structures. Using a ND-AMC and ND-graphite core-mantle component along with silicates and graphite very good extinction fits are obtained for the four non-CCM sites. Good fits are also obtained for other anomalous sightlines. The models use smaller proportion of silicates. The use of ND in the models is small and do not affect the carbon abundance constraints. That the ND component is generally covered in a mantle could explain the absence of diamond C--H stretching along most sightlines \citep{acke06}.

The grain growth and processing will in general lead to amorphous structures \citep{liGreenberg98}. Amorphous structures have both sp3 and sp2 carbons and the possibility of ND as a form of carbon matter can not be ruled out. ND must be considered particularly for sightlines with strong UV source and strong extinction. For further improvement distributions other than MRN need to be considered and strict abundance constraints \citep{clayton03a} applied. Availability of polarization data and its fitting can further constrain the component composition and shape effects.

\section*{Acknowledgments}

Use of library and computational facilities at IUCAA, Pune is acknowledged.

\clearpage


\begin{table*}
\centering
\caption{The models for the four non-CCM sightlines. Relative fraction of model components, $\chi^2$ and $(R_V)$ values are shown \label{tbl-1}}
\begin{tabular}{ccccccccccccc} \hline \hline
 & & \multicolumn{2}{c}{HD 210121} &   \multicolumn{2}{c}{HD 204827} &  \multicolumn{2}{c}{HD 29647} &   \multicolumn{2}{c}{HD 62542} \\

 &  &  \multicolumn{2}{c}{$R_V$ = 2.012}  &    \multicolumn{2}{c}{$R_V$ =  2.441} &   \multicolumn{2}{c}{$R_V$ = 3.456} &  \multicolumn{2}{c}{$R_V$ =  2.578}  \\


Model & Cut-off radii & Fraction & $\chi^2~(R_V) $  & Fraction & $\chi^2~(R_V) $ &   Fraction & $\chi^2~(R_V) $ &   Fraction& $\chi^2~(R_V) $ \\ \hline

Graphite     &  [0.001,0.120]  &  0.36  & 0.177 (1.43) & 0.40  & 0.102 (2.18)  &  0.36   & 0.253 (1.78) &   0.23   & 0.570 (1.93)  \\
Silicate     &  [0.001,0.200]  &  0.45\footnotemark[1]  &              &   0.65  & &   0.48   &              &   0.77   &   \\
\hline

ND-Graphite  & [0.001,0.120]   &  0.35  & 0.165 (1.41) &   0.40  & 0.114 (2.16) &   0.36   & 0.237 (1.77) &   0.23   & 0.564 (1.92) \\
Silicate     & [0.001,0.200]   &  0.45\footnotemark[1]  &               &  0.64  &              &   0.47   &              &   0.76   &   \\
\hline

Graphite     & [0.001,0.120]   &  0.18  & 0.027 (2.14) &   0.31  & 0.034 (2.15) &   0.23   & 0.202 (1.98) &   0.02   & 0.158 (2.24) \\
ND-AMC       & [0.001,0.150]   &  0.61  &              &   0.50  &              &   0.57   &              &   0.91   &   \\
Silicate     & [0.001,0.200]   &  0.18\footnotemark[1]  &              &   0.02  &              &   0.12   &              &   0.00   &   \\
\hline

Graphite     & [0.004,0.250]   &        &              & &        &               0.16    & 0.047 (2.87) & &         &   \\
ND-Graphite  & [0.001,0.120]   &        &              & &        &               0.17    &              & &         &   \\
ND-AMC       & [0.001,0.200]   &        &              & &        &               0.48    &              & &         &   \\
Silicate     & [0.001,0.200]   &        &              & &        &               0.00    &              & &         &   \\
\hline

Graphite     & [0.004,0.250]   &        &              & &        &               0.16    & 0.048 (2.85) & &         &   \\
ND-Graphite  & [0.001,0.120]   &        &              & &        &               0.21    &              & &         &   \\
ND-AMC       & [0.001,0.200]   &        &              & &        &               0.43    &              & &         &   \\
Silicate     & [0.001,0.200]   &        &              & &        &               0.02    &              & &         &   \\
\hline

Graphite     & [0.004,0.250]   &        &              & &        &              & &          0.04   & 0.1585 (2.23) \\
ND-AMC       & [0.001,0.150]   &        &              & &        &              & &          0.88   &   \\
Silicate     & [0.001,0.200]   &        &              & &        &              & &          0.02   &   \\

\hline
\end{tabular}
\end{table*}
\footnotesize{1~Silicate cut-off radii [0.001,0.120]}

\clearpage

\begin{table*}

\caption{For objects with steep FUV rise relative fraction of model components and $\chi^2$ and $R_V $ values are shown \label{tbl-2}}
\begin{tabular}{ccccccccccccc} \hline \hline

star & Model & Cut-off radii & Fraction & $\chi^2 $ &$R_V\textit{cal.} (R_V \textit{obs}) $ \\
spectral type & &  &  &  & \\\hline

HD 3191   & Graphite    &  [0.004,0.250]  &   0.40    & 0.0466  &  2.95(2.81) \\
B1 IV     & ND-AMC      &  [0.001,0.200]  &   0.29    &         &             \\
          & Silicate    &  [0.001,0.120]  &   0.15    &         &             \\
\hline

HD 13659  & Graphite    &  [0.004,0.250]  &   0.21    & 0.0195  &  2.60(2.49) \\
B1 Ib     & ND-AMC      &  [0.001,0.200]  &   0.45    &         &             \\
          & Silicate    &  [0.001,0.120]  &   0.12    &         &             \\
\hline

HD 17443  & Graphite    &  [0.004,0.250]  &   0.35    & 0.0200  &  2.35(2.58) \\
B9 V      & ND-AMC      &  [0.001,0.150]  &   0.58    &         &             \\
          & Silicate    &  [0.001,0.200]  &   0.02    &         &             \\
\hline

HD 30123  & Graphite    &  [0.004,0.250]  &   0.25    & 0.1181  &  2.81(3.30) \\
B8 III    & ND-AMC      &  [0.001,0.200]  &   0.70    &         &             \\
          & Silicate    &  [0.001,0.120]  &   0.00    &         &             \\
\hline

HD 284841 & Graphite    &  [0.004,0.250]  &   0.30    & 0.0493  &  2.96(2.98) \\
B9 II     & ND-AMC      &  [0.001,0.200]  &   0.65    &         &             \\
          & Silicate    &  [0.001,0.120]  &   0.04    &         &             \\
\hline

HD 287150 & Graphite    &  [0.004,0.250]  &   0.28    & 0.0459  &  3.20(3.29) \\
A2 V      & ND-AMC      &  [0.001,0.200]  &   0.54    &         &             \\
          & Silicate    &  [0.001,0.120]  &   0.10    &         &             \\
\hline
\end{tabular}
\end{table*}


\clearpage

\clearpage

\label{lastpage}

\end{document}